\documentclass[12pt]{iopart}

\usepackage{graphicx}

  \expandafter\let\csname equation*\endcsname\relax
  \expandafter\let\csname endequation*\endcsname\relax
\usepackage{amsmath}
\usepackage{url}

\graphicspath{{fig/}}

\usepackage{lineno}

\begin{document}


\title{The $\phi(1020)\rightarrow e^{+}e^{-}$ meson decay measured with the STAR experiment in Au+Au collisions at $\sqrt{s_{_{NN}}}$ = 200 GeV}

\author{
L.~Adamczyk$^{1}$,
J.~K.~Adkins$^{20}$,
G.~Agakishiev$^{18}$,
M.~M.~Aggarwal$^{31}$,
Z.~Ahammed$^{49}$,
I.~Alekseev$^{16}$,
A.~Aparin$^{18}$,
D.~Arkhipkin$^{3}$,
E.~C.~Aschenauer$^{3}$,
A.~Attri$^{31}$
% ...
}

\address{$^{1}$AGH University of Science and Technology, FPACS, Cracow 30-059, Poland}
% ...
\address{$^{3}$Brookhaven National Laboratory, Upton, New York 11973}
% ...
\address{$^{16}$Alikhanov Institute for Theoretical and Experimental Physics, Moscow 117218, Russia}
% ...
\address{$^{18}$Joint Institute for Nuclear Research, Dubna, 141 980, Russia}
% ...
\address{$^{20}$University of Kentucky, Lexington, Kentucky, 40506-0055}
% ...
\address{$^{31}$Panjab University, Chandigarh 160014, India}
% ...
\address{$^{49}$Variable Energy Cyclotron Centre, Kolkata 700064, India}
% ...

\date{\today}

\begin{abstract}
We report the measurement of the leptonic ($e^{+}e^{-}$) decay channel of the $\phi$(1020) meson
in Au+Au collisions at $\sqrt{s_{_{NN}}}$ = 200 GeV by the STAR experiment.
The transverse momentum ($p_{\rm T}$) spectrum is measured for 0.1 $\le p_{\rm T} \le 2.5$ GeV/$c$ at mid-rapidity ($|y|\le1$).
We obtain the $p_{\rm T}$-integrated $\phi$(1020) mass $M_{\phi}=1017.7\pm0.8 (\rm {stat.}) \pm0.9 (\rm {sys.})$ MeV/$c^{2}$ and width $\Gamma_{\phi} = 8.0\pm 2.5(\rm {stat.}) \pm 2.3(\rm {sys.}) \textrm{ MeV/}c^{2}$, which are within 1.5\,$\sigma$ and 1.1\,$\sigma$ of the vacuum values, respectively. 
No significant difference is observed in the measured $p_{\rm T}$ spectrum, $dN/dy$, or $\left<p_{\rm T}\right>$ of the $\phi$(1020) meson between the $e^{+}e^{-}$ and hadronic ($K^{+}K^{-}$) decay channels as measured by the same experiment. 
The experimental results are compared to a theoretical model including medium-modified $\phi$(1020).
\end{abstract}

\maketitle

%

\section{Introduction}
Experiments at Relativistic Heavy Ion Collider (RHIC)  have found a variety of indications for the deconfinement phase transition and formation of Quark-Gluon Plasma (QGP)~\cite{Adams2005}. 
Chirality, a fundamental symmetry of QCD which is spontaneously broken in vacuum, is predicted to be restored within an energy density range similar to the deconfinement phase transition~\cite{Karsch:2002aa}. The high energy density and temperature achieved in relativistic heavy-ion collisions provide a unique environment to study the chiral phase transition, which is of crucial importance in the Standard Model.
The properties of hadronic resonances, which are excitations of a (partially) chirally restored medium, are measured to study chiral symmetry restoration in hot and dense matter because short lifetimes allow a significant fraction of those produced to decay in-medium.

Among resonances, the $\phi$(1020) meson has several attractive features which can be used to study the two phase transitions.
The $\phi$(1020) meson, which is an $s\bar{s}$ bound state, is an ideal probe for the strangeness production in the medium~\cite{Rafelski1982}. 
As a vector meson, it can decay not only into kaons, but also into dileptons via a virtual photon state. 
Absence of strong interactions with final state hadrons in the dilepton channel permits reconstruction of $\phi$(1020) decays from throughout the collision evolution.
The mass, width, and branching ratios of $\phi$(1020) are predicted to be modified in a nuclear medium~\cite{PhysRevLett.66.2720,Hatsuda1992,Klingl:1998fk,Rapp1999a}.
Since the mass of $\phi$(1020) is just above the mass of two kaons, a decrease in the $\phi$(1020) mass may lead to a change in the relative production rates of $\phi(1020)\rightarrow e^{+}e^{-}$ and $\phi(1020)\rightarrow K^{+}K^{-}$ due to phase space limitations in the hadronic channel.
Even though the lifetime of $\phi$(1020) in vacuum (44 fm/$c$~\cite{PDG2014}) is longer than the estimated lifetime of the QGP (4-10 fm/$c$~\cite{PhysRevC.71.044906}), interactions with the medium may lead to an increase of the width, effectively shortening the lifetime.

\section{Analysis}
In this paper, we report the measurement of $\phi$(1020) production via the $e^{+}e^{-}$ decay channel in Au+Au collisions at $\sqrt{s_{_{NN}}} = 200$ GeV with the STAR experiment~\cite{Ackermann2003} at RHIC. 
The results presented in this paper are consistent with a recent STAR publication within systematical and statistical uncertainties, although used analysis methods differ~\cite{PhysRevC.92.024912}.
The detector subsystems used in this analysis are the Time Projection Chamber (TPC)~\cite{Anderson2003} and the Time-of-Flight (TOF)~\cite{Llope2012}.
\subsection{Event Selection}
Events are selected with collision vertices within $\pm$ 30 cm of the center of the TPC along the beam axis ($V_{z}$), and within 2 cm transverse of the beam axis to exclude interactions with the beam pipe. To suppress background from overlapping collisions in the TPC and ensure that the TOF start time is calculated for the triggered collision, the Vertex Position Detector (VPD)~\cite{Llope201423}, which provides the TOF start time, is used to select events for which $| V_{z,VPD} - V_{z,TPC} | < 3$ cm.
After an additional selection of the 0-80\% most central events (minimum bias for this analysis) based on the multiplicity of charged particles at mid-rapidity~\cite{Miller:2007fk}, about 250 million events are analyzed in this study. 

\subsection{Track Selection}
In order to reduce the contamination by daughter particles from weak decays (which typically have a decay length of several cm) and the $e^{\pm}$ tracks from $\gamma$ conversions in the material of the beam pipe, we require the distance of closest approach to the primary collision vertex of tracks to be less than 1.1 cm (where the resolution is typically 0.2-0.3 cm).
Tracks are required to have at least 22 points in the TPC.
The ratio of number of measured to possible TPC points in the TPC must be greater than 0.52 to avoid split tracks. 
The transverse momenta ($p_{\rm T}$) of the tracks have to be larger than 0.18 GeV/$c$ to reach the TOF detector, which is mounted outside of the TPC. 
The momenta of the electrons (including positrons if not specified) are required to be less than 2 GeV/$c$ due to the large identification contamination by charged hadrons above 2 GeV/$c$.

 \begin{figure}[h!bt]
 \centering 
 \includegraphics[width=0.9\linewidth]{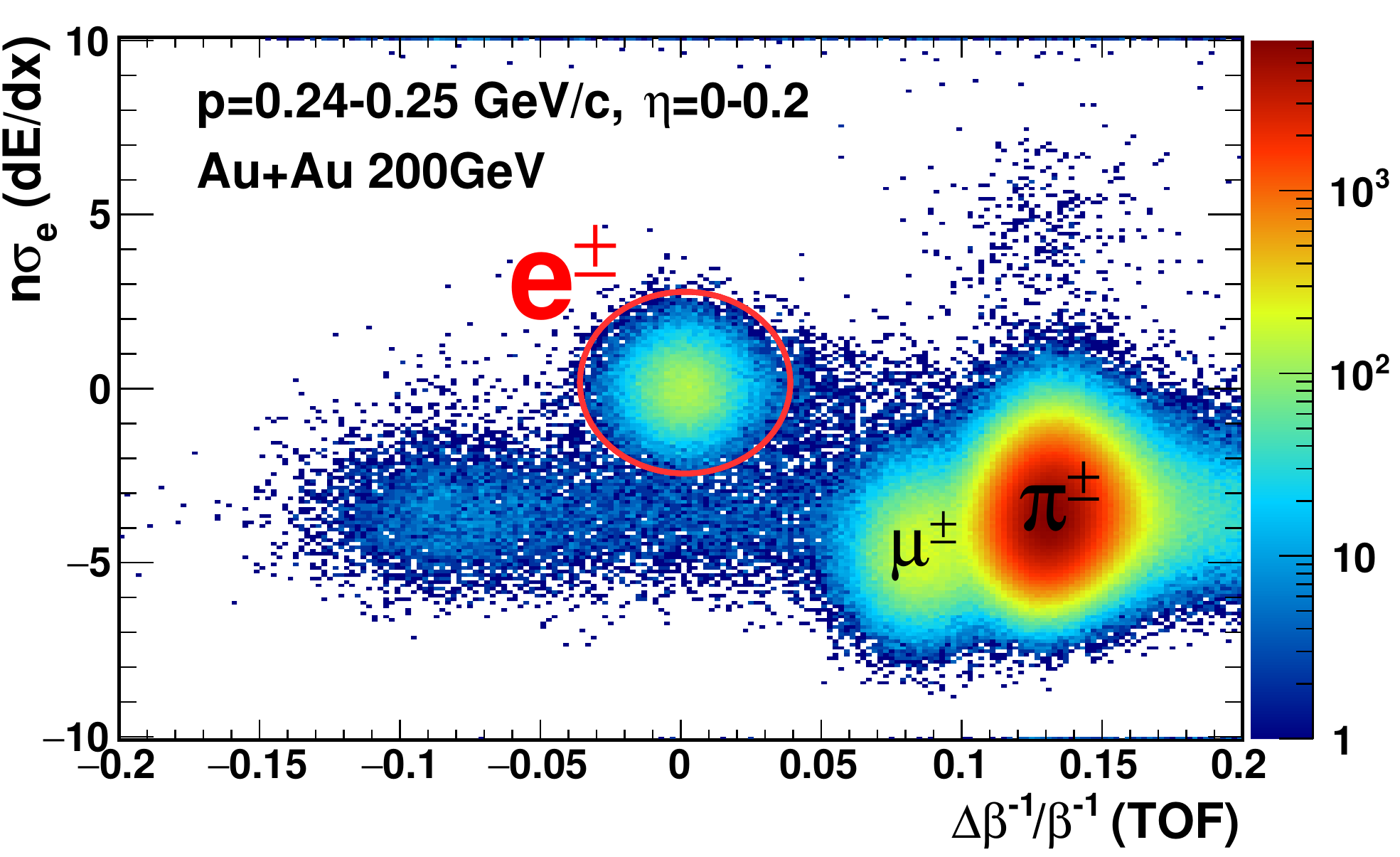}%
 \caption{(Color online) The distribution of $\text{n}\sigma\text{(dE/dx)}$ vs. $\Delta\beta^{-1}/\beta^{-1}$ for electrons, pions, and muons within a low momentum range. The red ellipse shows the cut to select electron candidates. \label{fig:pid2d}}
 \end{figure}
 \subsection{Electron Identification}
To achieve best separation of electrons from other particle species, the velocity measurement $\beta$ from the TOF and the ionization energy loss (dE/dx) measurement from the TPC are used simultaneously. We apply particle identification (PID) cuts in a two-dimensional (2D) space: $\text{n}\sigma\text{(dE/dx)}\equiv(X-\mu_{X})/\sigma_{X}\text{ with }X=\ln\text{(dE/dx)}$, which is a Gaussian distribution with mean $\mu_{X}$ and sigma $\sigma_{X}$, and $\Delta\beta^{-1}/\beta^{-1} = \left(\beta^{-1}_{TOF}-\beta^{-1}_{e}\right)/\beta^{-1}_{TOF}$, which is the deviation from electron expectation. 
 First, we select electrons requiring the distance from the mean of the electron distribution in the 2D space within 3$\sigma$ (the red ellipse in Fig.~\ref{fig:pid2d}), where $\sigma$ for both dE/dx and TOF depend on the momentum and pseudo-rapidity ($\eta$) of the measured particles.
In order to further suppress pion contamination, we apply an additional particle selection criteria based on the 2D probability~\cite{MasaW2013} that a particle is an electron, which is defined as
\begin{equation} \label{eq:prob}
   P^{e}(X, Y)\equiv\frac{N^{e}\times \textsc{Pdf}(X; \boldsymbol{\theta}^{e}_{X})\times \textsc{Pdf}(Y;  \boldsymbol{\theta}^{e}_{Y})}
   {\sum\limits_{i}N^{i}\times \textsc{Pdf}(X; \boldsymbol{\theta}^{i}_{X})\times \textsc{Pdf}(Y; \boldsymbol{\theta}^{i}_{Y})},
 \end{equation}
where $\textsc{Pdf}$ is a probability density function and $\boldsymbol{\theta}^{i}_{X,Y}$ are parameter sets for particle $i$. $N^{i}$ is the fraction of yield of particle $i$ ($\sum_{i}N^{i}=1$). All the parameters and the fraction of yields are obtained by fitting particle distributions in the 2D space. The summation in the denominator includes only electrons and pions for simplification in this analysis. All $\textsc{Pdf}$s are Gaussian functions except the $\Delta \beta^{-1}/ \beta^{-1}$ distribution for pions, for which the Student's $t$ function~\cite{bishop2006pattern} is used to better describe the tails of the distribution.
In this analysis, a probability larger than 60\% is required for a particle to be identified as an electron. 
The purity of the selected electrons is estimated to be about 95\% in $0.18 ~\text{GeV}/c \le p_{\rm T} \le 2.0~\text{GeV}/c$ by studying distributions of all relevant particle species such as pions, kaons and protons.   
 
\subsection{$\phi$(1020) Signal Extraction}
The invariant mass of $\phi$(1020) is reconstructed from all combinations of identified electron-positron pairs which have opening angles wider than 30 degrees in the same event (unlike-sign signal), shown as the black curve in the inset of Fig.~\ref{fig:fitContour}. 
The mixed-event technique~\cite{rho0_1,kstarAuAu}, which pairs electrons with positrons from 20 different events, is adopted to estimate the combinatorial background.
In order to preserve the event characteristics in the mixed-event technique, $e^{+}$ and $e^{-}$ pairs are selected only from the same event class, as defined by $V_{z}$ position and event plane angle~\cite{Snellings:2011uq} with ten bins each, and eight centrality bins. 
After normalizing the mixed-event background to the unlike-sign signal distribution in the invariant mass regions of
0.91-0.95 GeV/$c^{2}$ and 1.11-1.15 GeV/$c^{2}$
(orange areas in the inset of Fig.~\ref{fig:fitContour}), it is subtracted bin by bin from the unlike-sign signal distribution to extract the $\phi$(1020) signals. The resulting raw distribution for pairs with $|y|\le 1$ and $p_{\rm T}=0.1-2.5$ GeV/$c$ is shown in Fig.~\ref{fig:fitContour} as black points. 
A combined fit to the raw distribution, shown as the black curve in Fig.~\ref{fig:fitContour}, is performed using a Voigt function to describe the signal and a quadratic polynomial to represent the residual background.
The raw $\phi$(1020) yield is the sum of the counts in each bin above the residual background fit for the range $M_{\phi}-30 < M_{ee}<M_{\phi}+30 \textrm{ MeV/}c^{2}$ (which typically accounts for 85\% of the total yield) and the integral of the peak fit function outside that signal range.
We obtain a signal significance, defined as the raw signal yield over statistical error, of 15.9.
To measure the $p_{\rm T}$ spectrum we divide the integrated signal into seven $p_{\rm T}$ bins 
with a significance of about 6-7 in each $p_{\rm T}$ bin.

\subsection{Efficiency Correction for the $p_{\rm T}$ Spectrum}
The $p_{\rm T}$ spectrum is corrected for the TPC track reconstruction efficiency, the matching efficiency between TPC tracks and TOF hits, the detector acceptance, the track quality cut efficiency, and the PID cut efficiency~\cite{MasaW2013}. The track reconstruction efficiency and the TPC acceptance are obtained by embedding $\phi(1020)\rightarrow e^{+}+e^{-}$ tracks from Monte Carlo simulation into real Au+Au collision data at the detector response level and reconstructing them along with real tracks, such that intrinsic detector resolutions and inefficiencies are taken into account.
Along with the intrinsic resolutions from the simulation, we further smear the simulated electrons' momenta by $\sigma_{p_{\rm T}} = 0.51\%\,p_{\rm T}$ based on studies of the invariant mass line shape of $J/\psi$, where we do not expect any medium modification to the width.
The ratio of number of $\phi$(1020) mesons after quality cuts to number of simulated $\phi$(1020) mesons is taken as the efficiency of the reconstruction with quality cuts.
A sample of $e^{\pm}$ tracks from $\gamma$ conversions in the beam pipe (identified with 89\% purity) is used to estimate the TOF matching efficiency times acceptance.
The PID cut efficiency is calculated directly from the two-dimensional probability distributions mentioned above~\cite{MasaW2013}. 
The overall efficiency times acceptance is about 9\% for the lowest $p_{\rm T}$ bin and about 4\% for the highest $p_{\rm T}$ bin.

\subsection{Mass and Width Extraction}
To extract mass and width from invariant mass signals we use a Voigt function~\cite{Humlek1982437}, which is a Breit-Wigner function convolved with a Gaussian function accounting for the detector resolution, defined as 
\begin{equation} \label{eq:voigt}
   V(M;\sigma,\Gamma)\equiv\int_{-\infty}^{\infty} dM' G(M;M', \sigma) BW(M'; M_{0}, \Gamma),
 \end{equation}
where $G(M;M', \sigma)$ is a Gaussian function with mean $M'$ and sigma~$\sigma$, and $BW(M'; M_{0}, \Gamma)$ is a non-relativistic Breit-Wigner function with mass $M_{0}$ and width $\Gamma$.
To resolve degeneracy between the resolution parameter $\sigma$ and the $\phi(1020)$ width, a simulation with no intrinsic $\Gamma$ broadening is used to determine the value of $\sigma$ (i.e. $\Gamma$ can be fixed to the vacuum value~\cite{PDG2000}). It is found that 1.16 MeV/$c^{2}$ of additional $p_{\rm T}$ smearing beyond the intrinsic simulated resolution is necessary for a good fit, resulting in a value of 8.1 MeV/c$^{2}$ for $\sigma$, and further adding 0.04 MeV/c$^{2}$ to the extracted mass $M_{0}$.
This width extraction method is different from the recent STAR publication~\cite{PhysRevC.92.024912}, where the reported width includes the detector resolution. Taking the difference into account, the two results on the $\phi(1020)$ width are consistent within statistical and systematical errors.

\subsection{Systematical Errors Estimation}
The systematical errors are estimated by considering uncertainties due to: 
(1) mismatches between the real and simulated track distributions which lead to uncertainties in the track reconstruction efficiency,
(2) the PID selection criteria, mainly due to uncertainty in the 2D probability function,
(3) the TOF matching efficiency, and 
(4) variations due to the choice of the signal extraction method, such as the fit functions for the signal (non-relativistic \& relativistic Breit-Wigner function), the residual background function (first \& second order polynomials), and the normalization ranges for the mixed-event background.
These systematical uncertainties on yield, mass, and width are averaged over all $p_{\rm T}$ bins and summarized in Tab.~\ref{tab:sysSummary}.
The systematical uncertainty due to the additional $p_{\rm T}$ smearing is estimated as 0.2 MeV/$c$ (2.5\%) in width, which is insignificant within the statistical and other systematical errors of our measurement and not added in the results.
\begin{table}[ht]
 \centering
\begin{tabular}{ c  c  c  c }
\br
\textbf{Category} & \textbf{Yield} & \textbf{Mass} & \textbf{Width} \\ 
\mr
TPC track quality cut &	9.3\% & 0.083\% & 10.0\%\\ 
Particle identification (PID)  & 4.6\% & 0.043\% & 5.9\%\\
TOF matching efficiency & 8.3\% & 0.0 & 0.0 \\
Inv. mass signal extraction & 3.8\% & 0.007\% & 3.4\% \\
Additional $p_{\rm T}$ smearing & 0.0 & 0.0 & 2.5\% \\
\br
\end{tabular}
  \caption{Systematical uncertainties averaged over $p_{\rm T}$ bins, shown as percentages of the measured values.}
  \label{tab:sysSummary}
\end{table}

\section{Results}
\subsection{Mass and Decay Width}
\begin{figure}[h!bt]
\includegraphics[width=\linewidth]{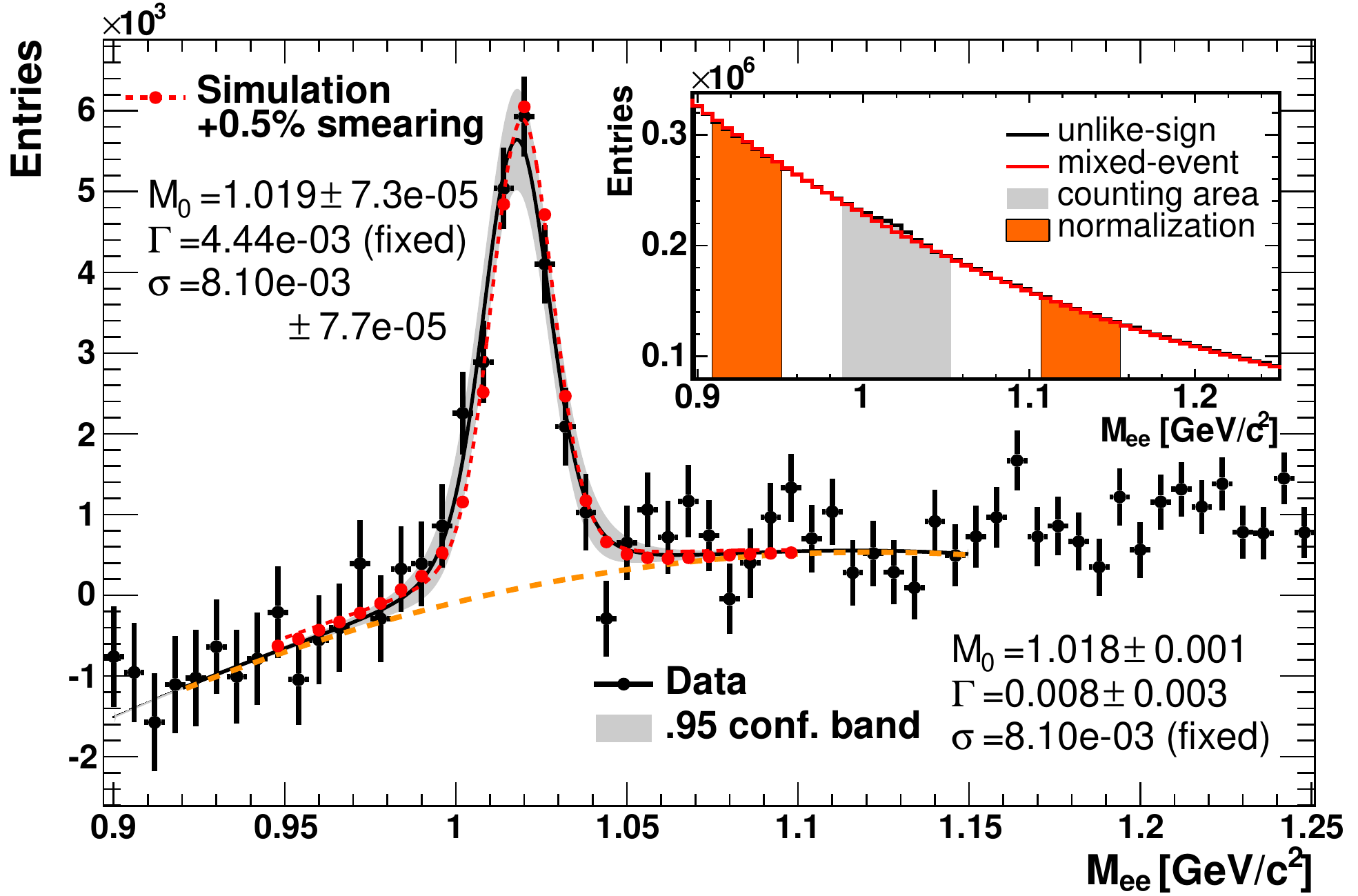}
\caption{(Color online) 
Black points are the $p_{\rm T}$-integrated $\phi(1020) \rightarrow e^{+}e^{-}$ signal from data after subtracting mixed-event background. 
The black curve is a fit to the signal with the Voigt function plus a quadratic polynomial, which represents residual background (orange long-dashed curve),
and the gray band is a 95\% confidence band of the fit. 
Red points and the short-dashed curve are the simulated $\phi(1020)$ signal with the additional momentum smearing and the fit function to the simulation, respectively, with the residual background from fitting the data added.
(Inset) Invariant mass distribution of the $e^{+} e^{-}$ pairs in the same event shown as a black curve, and normalized mixed-event background shown as a red curve. The orange areas show the normalization regions for the mixed-event background. Fit parameter units are GeV/$c^2$.
\label{fig:fitContour}}
\end{figure}
The $p_{\rm T}$-integrated $\phi$(1020) invariant mass distribution and fit result are shown in Fig.~\ref{fig:fitContour}.
The extracted fit parameters are $M_{\phi}=1017.7\pm0.8 (\textrm{stat.}) \pm0.9 (\textrm{sys.}) \textrm{ MeV}/c^{2}$ (1.5\,$\sigma$ away from the Particle Data Group (PDG) value of $1019.5\pm0.02\textrm{ MeV}/c^{2}$~\cite{PDG2014}) and $\Gamma_{\phi} = 8.0\pm 2.5(\rm {stat.}) \pm 2.3(\rm {sys.}) \textrm{ MeV}/c^{2}$ (1.1\,$\sigma$ away from the PDG value of $4.27\pm0.03\textrm{ MeV}/c^{2}$). 
We can extract an upper limit of 13.6 MeV/$c^{2}$ on the width and a lower limit of 1015.7 MeV/$c^{2}$ on the mass with a 95\% confidence level.

 \begin{figure}
 \centering
\includegraphics[width=\linewidth]{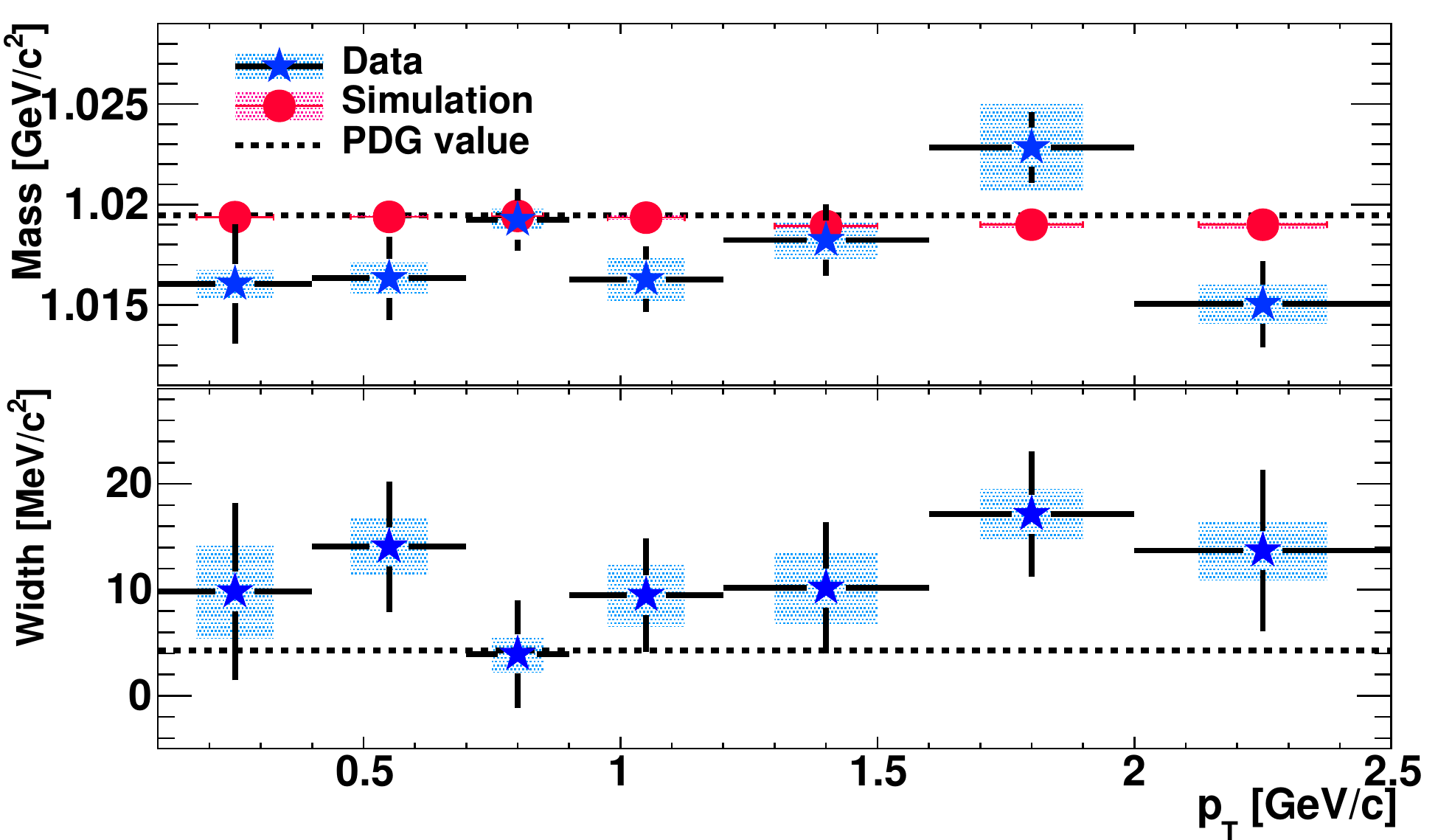}
 \caption{(Color online) Mass and width extracted from fitting as a function of $p_{\rm T}$. 
Dashed lines show the PDG $\phi$(1020) mass and width values.
Fit results to the data are shown as blue stars, whose statistical and systematical errors are represented by bars and filled boxes respectively. 
Red points are results of fits to the simulation with the additional momentum smearing (not shown for widths, as the fits use the fixed PDG value).
The systematical uncertainties on the width from the simulation are included in those of the real data.
 \label{fig:MassWidth}}
\end{figure}
Figure~\ref{fig:MassWidth} shows the extracted mass and width of $\phi$(1020) in each $p_{\rm T}$ bin from fitting the data.
The mass deviates by 1.4\,$\sigma$ and 1.6$\,\sigma$ for $p_{\rm T}=1.6-2$ GeV/$c$ and $2-2.5$ GeV/$c$ respectively from the simulation, which exhibits the expected vacuum values in the detector.
The deviations in width are 1.6\,$\sigma$ and 2.1\,$\sigma$ for $p_{\rm T}=0.4-0.7$ GeV/$c$ and $1.6-2$ GeV/$c$ respectively.
Most deviations are less than 2$\,\sigma$, but the measured masses are generally lower and the widths are systematically higher than vacuum values.
We do not observe any $p_{\rm T}$ dependence for mass and width within their errors.

\subsection{Comparison to Hadronic Decay Channel}
\begin{figure}[h!bt]
\includegraphics[width=\linewidth]{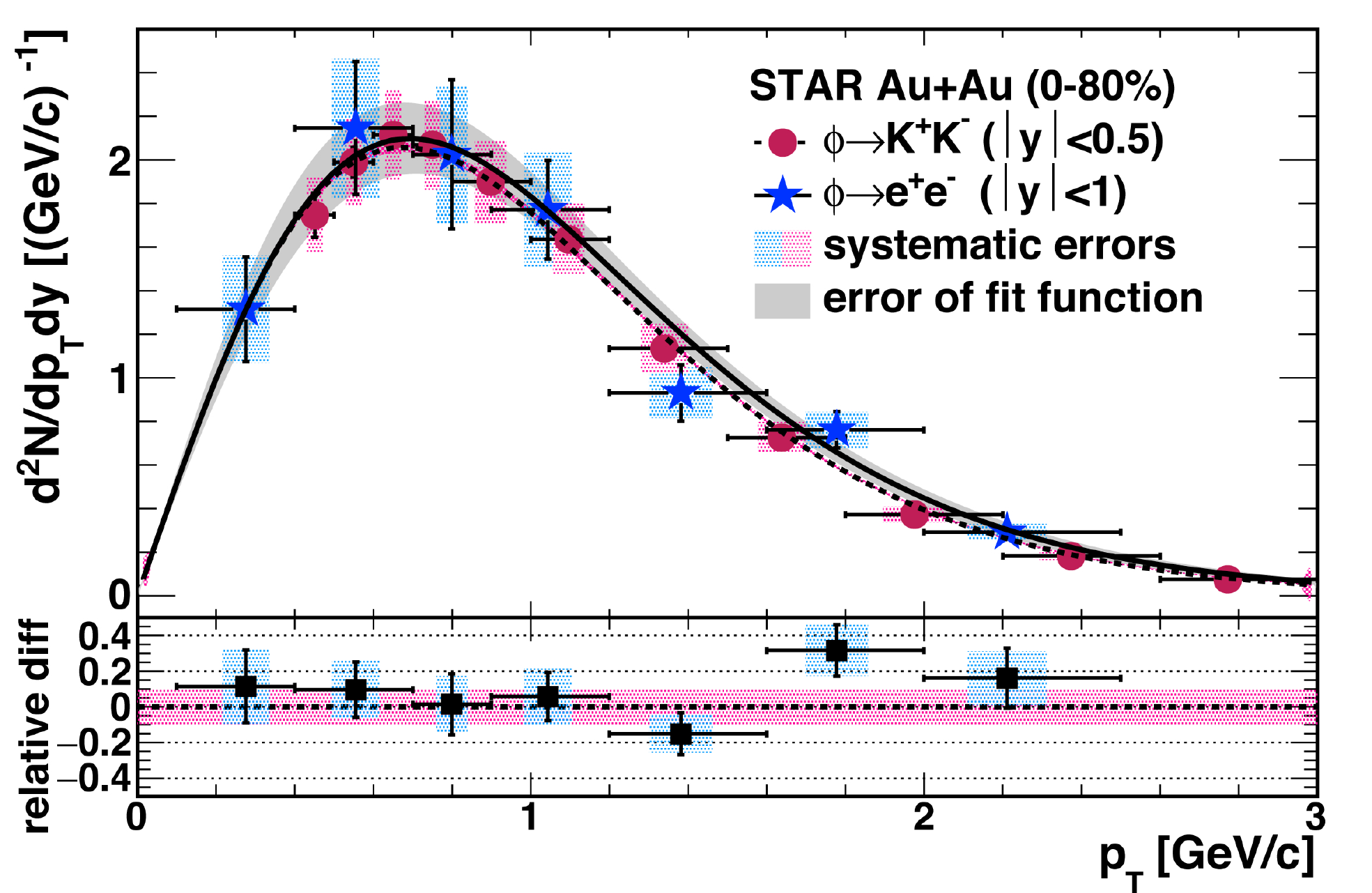}  
\caption{(Color online) 
The corrected $\phi(1020) \rightarrow e^{+}e^{-}$ (blue star) and $\phi(1020) \rightarrow K^{+}K^{-}$ (red circle)~\cite{Abelev2009} yields for the minimum bias events in Au + Au collisions at $\sqrt{s_{_{NN}}} = 200$ GeV. The vertical bars are statistical errors and the boxes are systematical uncertainties. Systematical errors for the $K^{+}K^{-}$ channel are fixed to be 10\% for all $p_{\rm T}$. The dashed and solid curves are $m_{\rm T}$-exponential fit functions to the hadronic and leptonic decay channel results, respectively. The gray band represents errors of the fit function to the leptonic result. Data points are placed at the mean $p_{\rm T}$ in each $p_{\rm T}$ bin estimated from the fit functions.
\label{fig:ptspectra}}
\end{figure}

The corrected $\phi(1020) \rightarrow e^{+}e^{-}$ invariant yields at rapidity $|y| \le 1$ for minimum bias Au+Au collisions at $\sqrt{s_{_{NN}}} = 200$ GeV are presented in Fig.~\ref{fig:ptspectra} along with $\phi(1020) \rightarrow K^{+}K^{-}$ results measured by STAR~\cite{Abelev2009} as well as $m_{\rm T}$-exponential fits~\cite{Abelev2009}. Both spectra are divided by the corresponding branching ratios to facilitate comparison. The inclusive $\phi$(1020) yield per unit rapidity $dN/dy$ is calculated by counting signals within the measured transverse momentum range and integrating the $m_{\rm T}$-exponential fit function for the rest. The mean transverse momentum $\left<p_{\rm T}\right>$ is obtained in the same way.  
The contributions from the unmeasured $p_{\rm T}$ regions, estimated via the extrapolation of the fitted function, are 3.7\% and 8.3\% for $dN/dy$ and $\left<p_{\rm T}\right>$, respectively.
We obtain $dN/dy$ = 2.91 $\pm$ 0.16(\rm {stat.}) $\pm$ 0.17(\rm {sys.}) and $\left< p_{\rm T}\right>$ = 1.03 $\pm$ 0.06(\rm {stat.}) $\pm$ 0.06(\rm {sys.}) GeV/$c$.
The values of $dN/dy$ and $\left<p_{\rm T}\right>$ are in agreement with the values for measurements from the hadronic decay channel ($dN/dy$ = 2.68 $\pm$ 0.15, $\left<p_{\rm T}\right>$ = 0.962 $\pm$ 0.015 GeV/$c$) within the errors (0.83\,$\sigma$ and 0.79\,$\sigma$ respectively).
In the bottom panel of Fig.~\ref{fig:ptspectra}, the relative differences between the two spectra, $(N_{ee}-N_{KK})/N_{KK}$, are shown. The $N_{KK}$ are calculated by integrating the fit function of the hadronic decay channel result (dashed curve) in each $p_{\rm T}$ bin.
The $p_{\rm T}$ spectrum of $\phi$(1020) measured in the $e^{+}e^{-}$ decay channel is in agreement with that from the hadronic decay within the statistical and systematical uncertainties. Similar results have been observed in the dimuon decay channel at SPS energies~\cite{Arnaldi2011325}.

\subsection{Comparison to a Theoretical Model}
We compare the measurements with a model calculation~\cite{Rapp:aa}, which is based on an effective chiral hadronic Lagrangian with coupling constants determined from the hadronic vacuum properties. 
This model includes an in-medium $\phi$ spectral function folded over the phase-space fireball evolution up to freeze-out~\cite{Rapp2001, VanHees2008}.
One of two components in the model is a ``cocktail'' contribution, which comes from the $\phi$(1020) decays after elastic hadronic interactions stop (kinetic freeze-out). 
The other component is from $\phi$(1020) decays in a thermal medium (thermal radiation) with a modified $\phi$(1020) spectral function. Its broadening implies back reactions regenerating $\phi$(1020) in the heat bath due to detailed balance.
In this model, only $\phi$(1020) decay after hadronization is included. 
Theoretically there can be an additional contribution, which is missing in this model, from medium-modified $\phi$(1020) decays in the QGP phase, if the $\phi$(1020) bound state can exist at temperatures above the critical temperature of the QGP.
 For this comparison, the STAR detector conditions (resolution, acceptance and efficiency) are applied to the model predictions.

\begin{figure}[h!bt]
\includegraphics[width=\linewidth]{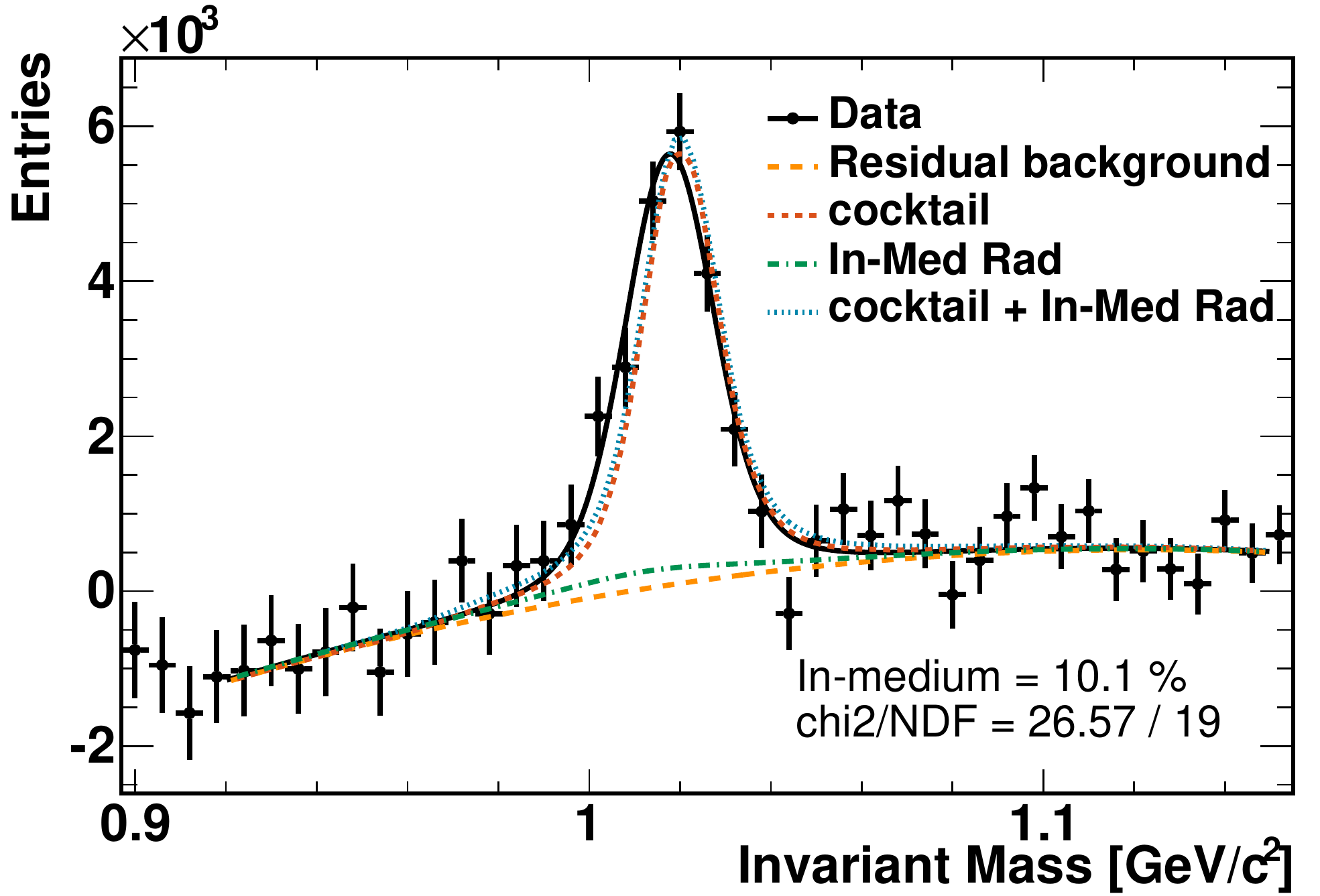}
\caption{(Color online) 
Model fit to the $\phi(1020)\rightarrow e^{+} e^{-}$ $p_{\rm T}$-integrated (over $p_{\rm T}=0.1-2.5$ GeV/$c$) invariant mass data, showing individual components of the model. The black curve is the Voigt fit function shown in Fig.~\ref{fig:fitContour}.
\label{fig:model}}
\end{figure}

The resulting invariant mass shape from the model is shown in Fig.~\ref{fig:model}. 
In the model calculation, the in-medium radiation component shows a mass shift of $\sim$7 MeV/$c^{2}$ and a width broadening of $\sim$55 MeV/$c^{2}$. 
Fitting the measured invariant mass distribution using the theoretical line shape, where the fraction of the in-medium contribution is fixed to be $10\%$, gives a $\chi^{2}/ndf$ of 1.40. 
However, by letting the fraction as a free parameter, the fit to data suggests a $29 \pm 15\%$ medium contribution with a $\chi^{2}/ndf=1.34$.

\section{Summary}
In summary, we have presented the measurements of $\phi$(1020) yield, mass, and width in the $e^{+}e^{-}$ decay channel at mid-rapidity ($|y|\le1$) in 0-80\% minimum bias Au+Au collisions at $\sqrt{s_{_{NN}}}$ = 200 GeV.
The measurements of $\phi$(1020) are consistent with its vacuum mass within 1.5~$\sigma$ and width within 1.1~$\sigma$ with no significant $p_{\rm T}$ dependence.
No significant difference is observed in the measured $\phi$(1020) $p_{\rm T}$ spectrum, $dN/dy$, or $\left<p_{\rm T}\right>$ between the leptonic ($e^{+}e^{-}$) and hadronic ($K^{+}K^{-}$) decay channels.
The measured invariant mass distribution is compared to a model calculation including a 10\% contribution from medium-modified $\phi$(1020) and a 90\% contribution from late decays in the vacuum. 
Our data demonstrate modest sensitivity to the medium-modified $\phi$(1020) component, but insufficient to observe definitively the small integrated effects predicted by this model.

\section{Acknowledgements}
\label{sec:acknowledgements}

We thank the RHIC Operations Group and RCF at BNL, the NERSC Center at LBNL, the KISTI Center in
Korea, and the Open Science Grid consortium for providing resources and support. This work was 
supported in part by the Office of Nuclear Physics within the U.S. DOE Office of Science,
the U.S. NSF, the Ministry of Education and Science of the Russian Federation, NSFC, CAS,
MoST and MoE of China, the National Research Foundation of Korea, NCKU (Taiwan), 
GA and MSMT of the Czech Republic, FIAS of Germany, DAE, DST, and UGC of India, the National
Science Centre of Poland, National Research Foundation, the Ministry of Science, Education and 
Sports of the Republic of Croatia, and RosAtom of Russia.
  
\section*{References}
\bibliography{AuAu200phi_iop}

\end{document}